\documentclass[aps, pra, reprint, amsmath, amssymb, superscriptaddress, showpacs]{revtex4-1}

\usepackage{CJK}
\usepackage{graphicx}
\usepackage{braket}
\usepackage{mathtools}
\usepackage{multirow}

\usepackage[usenames, dvipsnames]{color}
\usepackage{hyperref}
\hypersetup{colorlinks, citecolor=NavyBlue, linkcolor=ForestGreen, urlcolor=OrangeRed}
\usepackage{soul}

\begin{document}

\title{Theoretical characterization of the collective resonance states\\ underlying the xenon giant dipole resonance}

\begin{CJK*}{UTF8}{}

\author{Yi-Jen Chen \CJKfamily{bsmi}(陳怡蓁)}
\email{yi-jen.chen@cfel.de}
\affiliation{Center for Free-Electron Laser Science, DESY, Notkestrasse 85, 22607 Hamburg, Germany}
\affiliation{Department of Physics, University of Hamburg, Jungiusstrasse 9, 20355 Hamburg, Germany}

\author{Stefan Pabst}
\affiliation{Center for Free-Electron Laser Science, DESY, Notkestrasse 85, 22607 Hamburg, Germany}

\author{Antonia Karamatskou}
\affiliation{Center for Free-Electron Laser Science, DESY, Notkestrasse 85, 22607 Hamburg, Germany}
\affiliation{Department of Physics, University of Hamburg, Jungiusstrasse 9, 20355 Hamburg, Germany}

\author{Robin Santra}
\email{robin.santra@cfel.de}
\affiliation{Center for Free-Electron Laser Science, DESY, Notkestrasse 85, 22607 Hamburg, Germany}
\affiliation{Department of Physics, University of Hamburg, Jungiusstrasse 9, 20355 Hamburg, Germany}

\date{\today}

\begin{abstract}
We present a detailed theoretical characterization of the two fundamental collective resonances underlying the xenon giant dipole resonance (GDR). This is achieved consistently by two complementary methods implemented within the framework of the configuration-interaction singles (CIS) theory. The first method accesses the resonance states by diagonalizing the many-electron Hamiltonian using the smooth exterior complex scaling technique. The second method involves a new application of the Gabor analysis to wave-packet dynamics. We identify one resonance at an excitation energy of $74 \: \text{eV}$ with a lifetime of $27 \: \text{as}$, and the second at $107 \: \text{eV}$ with a lifetime of $11 \: \text{as}$. Our work provides a deeper understanding of the nature of the resonances associated with the GDR: a group of close-lying intrachannel resonances splits into two far-separated resonances through interchannel couplings involving the $4d$ electrons. The CIS approach allows a transparent interpretation of the two resonances as new collective modes. Due to the strong entanglement between the excited electron and the ionic core, the resonance wave functions are not dominated by any single particle-hole state. This gives rise to plasma-like collective oscillations of the $4d$ shell as a whole.
\end{abstract}


\maketitle

\end{CJK*}

\section{\label{sec:intro}Introduction}
The atomic xenon giant dipole resonance (GDR) has attracted much research interest since its discovery in 1964 \cite{d.ederer64a, a.lukirskii64a}, for it is one of the most prominent cases in atomic physics where many-body correlations play a conspicuous role. The GDR appears in the photoabsorption cross section of xenon as a pronounced and nearly symmetric hump centered around $100 \: \text{eV}$, with a width of about $40 \: \text{eV}$. The GDR lies in the electronic continuum above the $4d$ ionization threshold.

While the occurrence of the xenon GDR can be qualitatively explained by the independent-particle model \cite{j.cooper64a}, where a centrifugal barrier suppresses the $4d \rightarrow \epsilon f$ transitions near the $4d$ threshold, satisfactory agreement with experiments requires inclusion of many-body correlations beyond the mean-field level \cite{u.fano68a, m.amusia67a, w.brandt67a, a.starace70a, g.wendin73c}. Nowadays it is commonly accepted that the xenon GDR must be described as the result of the collective excitations of at least all the $4d$ electrons, forming short-lived plasma-like cooperative oscillations \cite{m.amusia00a, c.brechignac94a}. Because the GDR is a property of the inner-shell electrons, it is found in other atoms close to xenon in the periodic table and survives in molecules and solids \cite{m.amusia00a, c.brechignac94a}. Similar giant resonances also prevail in nuclei, metallic clusters, fullerences, etc \cite{m.amusia00a, c.brechignac94a}. 

A considerable number of measurements have been performed for a precise characterization of the xenon photoabsorption spectrum in the XUV with perturbative light sources \cite{r.haensel69a, j.west78a, u.becker89a, j.samson02a}. However, with the birth of various new source technologies, the old spectroscopic feature of the xenon GDR continues to enthrall start-of-the-art experiments. For example, high-harmonic generation (HHG) spectra of xenon driven by an intense mid-infrared laser display a striking enhancement in the plateau \cite{a.shiner11a}, which reflects the partial cross section of the $5p$ valence shell strongly modified by the GDR \cite{s.pabst13b}. Also, the GDR lies at the heart of the behavior of xenon exposed to free-electron lasers with ultrahigh XUV irradiance \cite{m.richter09a, v.richardson10a, n.gerken14a}. Hence, it is important to fully understand the nature of the xenon GDR.

Following the earliest independent-particle model \cite{j.cooper64a}, various advanced many-body theories have succeeded in reproducing the experimental cross section associated with the xenon GDR remarkably well  \cite{m.amusia90a, g.wendin73c, z.crljen87a, z.altun88a, a.zangwill80a}. Nevertheless, a fundamental question frequently overlooked is what exactly are the basic collective modes that give rise to the spectral properties of the xenon GDR. A work by Wendin in 1971 \cite{g.wendin71a} (see Ref.~\cite{g.wendin73c} for details) using the random-phase approximation with exchange (RPAE) identifies two collective resonances in this energy range. Another calculation by Lundqvist in 1980 \cite{s.lundqvist80a} utilizing a hydrodynamic treatment of electron density oscillations also finds two collective modes, but one of them sits at an energy incompatible with experimental observations. In addition to the very limited theoretical predictions of the resonance positions, neither of these studies explicitly specify the resonance widths. Consequently, the nature of the inherent collective resonances hidden in the broad spectral blur of the xenon GDR still remains an unsolved question.

The purpose of this paper is to provide a thorough characterization of the resonance substructures underlying the xenon GDR within the framework of the configuration-interaction singles (CIS) approach \cite{n.rohringer06a, l.greenman10a}, an \textit{ab initio} theory that can capture essential many-body effects in light-matter interactions \cite{s.pabst13a} including the xenon GDR \cite{d.krebs14a}. We resolve two collective dipolar resonances residing in the spectral range of the GDR, with one position differing from that given by Wendin \cite{g.wendin73c} by $15 \: \text{eV}$. Whereas Wendin resorted to an approximate condition only applicable to weakly damped plasma \cite{m.amusia74a} to estimate the positions of the collective excitations, this work provides the first quantitative results for the resonance positions and lifetimes. In contrast to the conventional view that many-body correlations only quantitatively shift and flatten the resonance as seen in the photoabsorption spectra \cite{j.cooper64a, u.fano68a, m.amusia67a, w.brandt67a, a.starace70a, a.zangwill80a}, we clearly demonstrate that many-body correlations qualitatively change the nature of the xenon GDR: a group of intrachannel resonances splits into two far-separated resonances as soon as we switch on interchannel interactions involving the $4d$ electrons. Since the resonance lifetimes are very short, the resonances strongly overlap and appear as one big hump in the photoabsroption cross section \cite{g.wendin73c}. In contrast to the plasma-type treatments used in Refs.~\cite{g.wendin73c, s.lundqvist80a}, the full many-body wave functions are directly obtained through our CIS approach. As the wave functions of the two resonances cannot be expressed by any single particle-hole state, we concretely show that they are indeed new collective modes \cite{g.wendin73c, m.amusia00a, c.brechignac94a}.

In this work, the isolation of the resonance substructures is consistently accomplished by means of two complementary and general methods implemented using CIS. The first, time-independent approach provides a comprehensive characterization of all the resonance properties by directly diagonalizing the many-body Hamiltonian using the smooth exterior complex scaling (SES) technique \cite{n.moiseyev98a, h.karlsson98a, c.buth07b}. Complex scaling \cite{n.moiseyev98b} has been used to solve the electronic resonance problem for few-electron atoms \cite{n.moiseyev79a, a.scrinzi98a, d.telnov02a, c.mccurdy02a} and molecules \cite{x.bian11a}. However, it has not yet been used to address collective resonances in many-electron atoms. The second, time-dependent approach involves a new application of the time-frequency Gabor analysis \cite{b.boashash03a, p.antoine95a} to the autocorrelation function of a wave packet. It is a common routine in molecular dynamics to look for resonance energies in the Fourier domain \cite{r.schinke93a, s.gray92a, a.isele94a}. Nonetheless, our analysis in the combined time-frequency domain not only shows improved performance in disentangling strongly overlapping resonances, but also supplies an appealingly intuitive view on the time evolutions of various wave-packet components.

The remainder of this article is structured as follows: Sec.~\ref{sec:theory} presents the theoretical tools. Sec.~\ref{subsec:cis} first lays the foundation of our many-body CIS scheme. Sec.~\ref{subsec:ses} and Sec.~\ref{subsec:gb} explain, respectively, the time-independent SES and time-dependent Gabor procedures to access multiple resonances. In Sec.~\ref{sec:result} we apply the methods to the xenon GDR, with the computational details in Sec.~\ref{subsec:comp}. The results of the SES and Gabor approaches are discussed separately in Secs.~\ref{subsec:datases} and \ref{subsec:datagb}. Restrictions imposed on the electronic-configuration space in Secs.~\ref{subsec:datases} and \ref{subsec:datagb} are justified in Sec.~\ref{subsec:5s5p}. Finally, Sec.~\ref{sec:conclude} concludes the study with a future outlook. Further numerical evidence indicating the consequence of finding resonance poles with the approximate condition used by Wendin \cite{g.wendin73c} is provided in the Appendix{\ref{app:dielec}.

Atomic units (a.u.) are used throughout the paper ($|e| = m_e = \hbar = 4 \pi \epsilon_0=1$) unless otherwise stated.

\section{\label{sec:theory}Theoretical methods}
\subsection{\label{subsec:cis}CIS theory}
In this work, the many-electron Schr\"{o}dinger equation is treated within the CIS framework, an \textit{ab initio} theory that allows one to encapsulate essential many-body physics beyond the mean-field Hartree-Fock picture \cite{s.pabst13a, a.szabo96a}. Our implementation of the CIS method has been successfully applied to a wealth of physical phenomena of many-electron atomic systems interacting with light fields \cite{s.pabst13a}, including perturbative \cite{s.pabst11a, d.krebs14a, e.heinrich-josties14a} and nonperturbative \cite{s.pabst13b, s.pabst14a, s.pabst12c} multiphoton processes with photon energies from the x-ray regime down to the near-infrared regime. Particularly, the ability of CIS to reproduce important features of the experimentally observed xenon GDR is demonstrated in Ref.~\cite{d.krebs14a}. In the following, we outline the formulation of our CIS approach. Further details can be found in previous publications \cite{n.rohringer06a, l.greenman10a, s.pabst12b}.
 
The nonrelativistic Hamiltonian for an $N$-electron atom in the absence of external fields can be generally written as
\begin{align}
\hat{H} &= \sum_{n=1}^N \left( \frac{\hat{\mathbf{p}}_n^2}{2} - \frac{Z}{\left| \hat{\mathbf{r}}_n \right|} + \hat{V}^{\text{MF}} (\hat{\mathbf{r}}_n) \right) - E^{\text{MF}}_0  \nonumber \\
&+ \bigg( \frac{1}{2} \sum_{\substack{n, n'=1 \\ n \neq n' }}^{N} \frac{1}{\left| \hat{\mathbf{r}}_n - \hat{\mathbf{r}}_{n'} \right|} - \sum_{n=1}^N \hat{V}^{\text{MF}}(\hat{\mathbf{r}}_n) \bigg) %
\nonumber \\
& \eqqcolon \hat{H}_0 + \hat{H}_1 \label{eq:hamiltonian}
\end{align}
where $\hat{\mathbf{p}}_n$ and $\hat{\mathbf{r}}_n$ are the momentum and coordinate operators for individual electrons, $Z$ is the nuclear charge, and $\hat{V}^{\text{MF}}$ is the mean-field potential contributing to the standard Fock operator \cite{a.szabo96a}. The Hartree-Fock ground state energy $E^{\text{MF}}_0$ is introduced to shift the entire energy spectrum for cosmetic purposes. The total Hamiltonian is divided such that $\hat{H}_0$ is merely a one-body operator and that all the residual two-body electron-electron Coulomb interactions beyond the description of the mean-field potential are contained in $\hat{H}_1$. 

The $N$-electron Hamiltonian is represented in the $N$-electron CIS configuration space:
\begin{equation}
\mathcal{V}_{\text{CIS}} \eqqcolon \left\{ \ket{\Phi^{\text{MF}}_0},  \ket{\Phi_i^a} \right\}, \label{eq:CISspace}
\end{equation}
which gives an ansatz for an $N$-electron wave function:
\begin{equation}
\ket{\Psi} = \alpha_0 \ket{\Phi^{\text{MF}}_0} + \sum_{i,a} \alpha_i^a \ket{\Phi_i^a}. \label{eq:ansatz}
\end{equation}
Thus, the Hilbert space is truncated and only consists of the Hartree-Fock ground state $\ket{\Phi^{\text{MF}}_0}$ plus its singly excited configurations $\ket{\Phi_i^a} = \hat{c}_a^\dagger \hat{c}_i \ket{\Phi^{\text{MF}}_0}$, with $\hat{c}_i$ annihilating an electron from an initially occupied orbital $i$ and $\hat{c}_a^\dagger $ putting it into an initially unoccupied orbital $a$ \cite{a.szabo96a}. The range of the index $i$ selects the active occupied orbitals from which an electron can be excited or ionized, i.e.~the accessible \textit{channels} \cite{a.starace82a}, thus enabling one to test the multichannel character of the overall physical process \cite{e.heinrich-josties14a, s.pabst13b, s.pabst12b}.

The matrix of the $N$-electron Hamiltonian is then either diagonalized (Sec.~\ref{subsec:ses}) or used in the time-dependent Schr\"{o}dinger equation (Sec.~\ref{subsec:gb}). In CIS, the only matrix elements that can lead to two-body effects are $\bra{\Phi_i^a} \hat{H}_1 \ket{\Phi_j^b}$. Specifically, it is the type of matrix elements with the indices $a \neq b$ and $i \neq j$, named \textit{interchannel-coupling} terms \cite{a.starace82a}, that permits the simultaneous change of the state of the excited electron and that of the ionic core, i.e., interchannel coupling leads to the formation of a correlated particle-hole pair \cite{s.pabst11a}. Numerically, we can tailor the two-body nature of $\hat{H}_1$ and study its influences by enforcing all interchannel-coupling matrix elements to be zero and considering only the matrix elements $\bra{\Phi_i^a} \hat{H}_1 \ket{\Phi_j^b}$ with $i = j$. In this scenario, called \textit{intrachannel-coupling} model \cite{a.starace82a}, $\hat{H}_1$ effectively acts as a one-body operator: once the electron is excited, it can sense the potential produced by the parent ion but cannot modify the ionic state, which is therefore forbidden to partake in many-body correlations \cite{s.pabst11a, s.pabst13b, d.krebs14a, e.heinrich-josties14a}.

\subsection{\label{subsec:ses}Time-independent approach to resonances:\\ SES method}
A conventional procedure to access eigenstates is the direct diagonalization of a general Hamiltonian. Nevertheless, being exponentially divergent in the asymptotic region renders resonance states, also known as Siegert \cite{a.siegert39a} or Gamow \cite{g.gamow28a} vectors, inadmissible elements of the Hilbert space of a Hermitian Hamiltonian. Standard techniques such as complex scaling \cite{n.moiseyev98b, y.ho83a, w.reinhardt82a} and the use of complex absorbing potentials (CAPs) \cite{r.santra02a, u.riss93a, u.riss98a} were thus developed to transform the wave function of a resonance state into a single square-integrable function. In this paper, we adopt the SES method \cite{n.moiseyev98b, h.karlsson98a, n.moiseyev98a, c.buth07b}, a variant of the complex scaling technique.

Our use of SES \cite{c.buth07b, s.pabst14b} relies on an analytic continuation of the radial part of the electron coordinate into the complex plane $r \rightarrow \rho (r)$ following the path of Moiseyev \cite{n.moiseyev98a} in the form of Karlsson \cite{h.karlsson98a} adapted to the purely radial problem presented here:
\begin{equation}
\rho(r) = r + (e^{i\theta}-1) \left[ r + \lambda \ln\left( \frac{1+e^{(r_0-r)/\lambda}}{1+e^{r_0/\lambda}} \right)  \right].  \label{eq:ses}
\end{equation} 
This path smoothly (depending on the parameter $\lambda$) rotates the electron radial coordinate for $r > r_0$ about an angle $\theta$ into the upper complex plane. 

Solving the eigenvalue problem of the complex-scaled Hamiltonian with the basis set $\mathcal{V}_{\text{CIS}}$, the resonance states can be uniquely identified as the exposed poles situated above the rotated continua in the complex-valued energy spectrum \cite{n.moiseyev98b, y.ho83a, w.reinhardt82a}. It is then straightforward to obtain the complex resonance energy or the Siegert energy \cite{a.siegert39a, r.santra02a, n.moiseyev98b}
\begin{equation}
E_{n} = \Xi_{n} - i\Gamma_{n}/2, \:\:\:  \Xi_{n}, \Gamma_{n} \in \mathbb{R}^+ \label{eq:energy}
\end{equation}
as well as the wave function $\ket{\Phi_{n}}$ associated with the $n$-th resonance state. In Eq.~(\ref{eq:energy}), $\Xi_{n}$ is the resonance position, and $\Gamma_{n}$ gives the inverse lifetime for the quasibound state to escape to the continuum. The detailed implementation of SES within CIS using a generalized finite-element discrete variable representation (FEM-DVR) will be addressed in a forthcoming publication \cite{s.pabst14b}.

The SES method shares with CAPs \cite{r.santra02a} the merit that it leaves the interior $r < r_0$ untouched and does not perturb the Hartree-Fock ground state (if $r_0$ and $\lambda$ are chosen suitably) \cite{h.karlsson98a, n.moiseyev98b}. At the same time, it eliminates many drawbacks of CAPs: no optimization with respect to a parameter is required in order to identify the resonance energies, and the whole transformation rests on the rigorous mathematical theory of complex scaling \cite{c.buth07b, h.karlsson98a, n.moiseyev98b}.

Notice that the complex-scaled Hamiltonian is no longer a Hermitian but a complex symmetric matrix \cite{c.buth07b}. As a result, the symmetric inner product $\left( \cdot \right| , \left| \cdot \right)$ must be used instead of the Hermitian inner product $\bra{\cdot} , \ket{\cdot}$ to ensure orthogonality relations \cite{l.greenman10a, r.santra02a, n.moiseyev98b}.

\subsection{\label{subsec:gb}Time-dependent approach to resonances:\\ Gabor analysis of autocorrelation functions}
Decoding the resonance substructures for a general quantum system can also be done through wave-packet propagation. The key physical quantity employed throughout our analysis is the time-dependent autocorrelation function defined as
\begin{equation}
C(t) \eqqcolon \left( \Psi(0) | \Psi(t) \right), \label{eq:autodef}
\end{equation}
where $\left.| \Psi(0) \right)$ is an initial state and $\left.| \Psi(t) \right)$ is the freely evolved wave packet at a later time $t$. Note that the symmetric inner product is adopted here. This is because in the time-dependent case we continue using SES, which effectively functions as a CAP and absorbs the outgoing flux reaching the end of the numerical grid \cite{h.karlsson98a, n.moiseyev98a, u.riss98a}. For an initial state $\left.| \Psi(0) \right)$ orthogonal to $|\Phi^{\text{MF}}_0)$, the time evolution of $\left.| \Psi(t) \right)$ is governed by the CIS coefficients $\alpha_i^a(t)$ [cf.~Eq.~(\ref{eq:ansatz})]. Inserting Eq.~(\ref{eq:ansatz}) into the time-dependent Schr\"{o}dinger equation with the complex-scaled Hamiltonian, one can derive and numerically integrate the equations of motion for $\alpha_i^a(t)$ \cite{l.greenman10a}:
\begin{equation}
i \: \dot{\alpha_i^a}(t) = \left( \Phi_i^a \right| \hat{H}_0 \left| \Phi_i^a \right) \alpha_i^a (t) %
+ \sum_{j,b} \left( \Phi_i^a \right| \hat{H}_1 | \Phi_j^b ) \: \alpha_j^b (t). \label{eq:eom}
\end{equation}

For a quantitative determination of the resonance energies, we assume that, by proper preparation, the initial state is essentially composed of the resonance states of interest and all the contributions from the bound states and the continuum can be ignored. Expanding $ \left| \Psi(0) \right)$ in terms of the orthonormal resonance wave functions $\left| \Psi(0) \right) = \sum_n a_n \left.| \Phi _n \right)$, Eq.~(\ref{eq:autodef}) then bears the following structure:
\begin{equation}
C(t) = \sum_n a_n^2 e^{-i \Xi_n t - \frac{\Gamma_n}{2} t } \label{eq:auto}.
\end{equation}
The validity of Eq.~(\ref{eq:auto}) and the resonances that can be extracted evidently depend on the quality of $\left.| \Psi(0) \right)$. How we prepare the wave packet ideal for studying the xenon GDR will be discussed in Sec.~\ref{subsec:datagb}.

A common strategy to infer Siegert energies from wave-packet propagation is to conduct a Fourier analysis and study the autocorrelation function in the frequency domain \cite{r.schinke93a, s.gray92a, a.isele94a}. Performing a one-sided Fourier transformation on Eq.~(\ref{eq:auto}) [assuming $C(t) = 0$ for $t < 0$] yields
\begin{align}
C(\omega) &= \frac{1}{\sqrt{2\pi}} \int_{0}^{+\infty} dt \:  e^{i \omega t} C(t) \nonumber \\
&= \sum_n \frac{a_n^2}{\sqrt{2\pi}} \frac{ \frac{\Gamma_n}{2} + i (\omega - \Xi_n) }{\frac{\Gamma_n^2}{4} + (\omega - \Xi_n)^2 }, \label{eq:ftauto}
\end{align}
i.e., a superposition of Lorentzian functions and dispersive curves parametrized by the Siegert energies \footnote{Notice that $a_n^2$ is a complex number and can impart additional phase to the following functions. This also results in the fact that it remains difficult to devise an analytic signal extending $C(t)$ into $t \in \left( - \infty, 0\right]$.}.

For a single resonance, the spectral distribution of the autocorrelation function reads
\begin{equation}
\left| C^{(1)}(\omega) \right|^2 = \frac{\left| a_1^2 \right|^2}{2 \pi} \frac{1}{\frac{\Gamma_1^2}{4} + (\omega - \Xi_1)^2}, \label{eq:ftsingle}
\end{equation}
which is a Lorentzian with a peak at $\Xi_1$ and a width of $\Gamma_1$. If more than one resonance exists, $\left| C(\omega) \right|^2$  comprises several Lorentzians and their interferences. Upon empirically specifying the number of resonance states, it is possible to retrieve the Siegert energies by numerically fitting $\left| C(\omega) \right|^2$ based on Eq.~(\ref{eq:ftauto}).

Next, we extend the standard spectral analysis to a time-frequency analysis \cite{b.boashash03a, p.antoine95a, x.tong00a} of the autocorrelation function and examine its information content in the combined \textit{time-frequency} domain. Applying a Gabor transformation \cite{d.gabor46a, p.antoine95a, c.chirila10a} to Eq.~(\ref{eq:auto}), we can derive
\begin{align}
&C_t(\omega) = \frac{1}{\sigma \sqrt{2 \pi}} \int_{0}^{+\infty} dt' \: e^{i\omega t'} e^{- \frac{(t'-t)^2}{2\sigma^2}} C(t') \label{eq:gb} \\
& \approx \sum_n a_n^2 e^{\frac{\sigma^2\Gamma_n^2}{8} - \frac{\Gamma_n}{2}t - \frac{\sigma^2}{2}(\omega - \Xi_n)^2 + i \sigma^2 (\frac{t}{\sigma^2} - \frac{\Gamma_n}{2})(\omega - \Xi_n) }.  \label{eq:gbauto}
\end{align}
Eq.~(\ref{eq:gb}) can be interpreted as gating the time-dependent signal by a Gaussian window function of width $\sigma$ centered at $t$. Due to the finite size of the window function and the sudden turn-on of the autocorrelation function at $t = 0$, the analytical expression in Eq.~(\ref{eq:gbauto}) works as a reasonable approximation to $C_t(\omega)$ when $t \gg \sigma$.

For a single resonance, the transient spectral distribution of the autocorrelation function at time $t$ can be simply approximated by
\begin{equation}
\left| C_t^{(1)}(\omega) \right|^2 \approx 
\left| a_1^2 \right|^2 e^{\frac{\sigma^2 \Gamma_1^2}{4}} e^{-\Gamma_1 t} e^{-\sigma^2(\omega - \Xi_1)^2},\label{eq:gbsingle}
\end{equation}
i.e.~a Gaussian with a peak at $\Xi_1$ and a width determined by $\sigma$. From the decay rate of the amplitude, $\Gamma_1$ can be extracted.

The advantage of the Gabor analysis over the Fourier analysis becomes apparent when multiple resonances come into play. In this situation, $\left| C(\omega) \right|^2$ comprises several Gaussians and their interferences. Consider the example where there are overlapping broad resonances yet with different lifetimes. Compared to the static information conveyed by the Fourier spectrum, it is more likely to detect the resonances through the time evolution of the Gabor profile, where their competition causes dynamics in the frequency distribution. Quantification of the resonance energies can be done by fitting $\left| C_t(\omega) \right|^2$ with the help of Eq.~(\ref{eq:gbauto}) at different time steps.

It is worthwhile to note that $C(\omega)$ is related to a measurable physical quantity---the photoabsorption cross section \cite{d.krebs14a, x.tong10a}---although we usually consider its modulus squared to 
reduce the number of irrelevant fitting parameters (e.g.~the phase of $a_n^2$). Also, a real physical observable---the dipole moment---can be used in the time-frequency analysis as well \footnote{The reason for analyzing the autocorrelation function instead of the dipole moment is because the former is a complex quantity, which often gives rise to neater analytical expressions for the (time-)frequency distributions.}. In this case, its frequency distribution is associated with the spectrum of electromagnetic radiation emitted by the system \cite{j.krause92a}. Lastly, it is tempting to point out the conceptual similarity between Gabor transformation and the spectrogram measured in a pump-probe experiment \cite{a.wirth11a, a.kaldun14a, e.power10a}, albeit there is no real probe pulse involved in the current theory.

\section{\label{sec:result}Results and discussion}
\subsection{\label{subsec:comp}Computational details}
The theoretical methods described in the previous Sec.~\ref{sec:theory} shall now be applied to the detailed resonance structure of the xenon GDR that is probed by linear spectroscopy using linearly polarized XUV light \cite{d.ederer64a, a.lukirskii64a, r.haensel69a, j.west78a, u.becker89a, j.samson02a}. The calculations are done using our \textsc{XCID} package \cite{xcid}. A single set of numerical parameters is employed throughout our calculations to compare consistently the results obtained by the two approaches.

Exploiting symmetries, $nl_{\pm m}$ is counted as one ionization channel \cite{n.rohringer06a, s.pabst12b}. In the energy range of concern, it is adequate to assume that electron depopulations only happen from the $4d$, $5s$, and $5p$ orbitals \footnote{The experimental binding energies for the $4d$, $5s$, and $5p$ electrons are $67.5 \: \text{eV}$ \cite{a.thompson09a}, $23.4 \: \text{eV}$ \cite{a.kramida14a}, and $12.1 \: \text{eV}$ \cite{a.kramida14a}, respectively. The value corresponding to a higher total angular momentum quantum number $j$ is taken, since its ionization threshold is lower. The average decay width of the $4d$ holes is $107.5 \: \text{meV}$ \cite{p.lablanquie02a}.}. However, for a meaningful comparison with the work by Wendin \cite{g.wendin73c}, the calculations presented in Secs.~\ref{subsec:datases} and \ref{subsec:datagb} are performed without activating the outer $5s$ and $5p$ shells. As we will see in Sec.~\ref{subsec:5s5p}, these channels only cause minor quantitative modifications. The HF orbital energies are slightly adjusted to match the experimental binding energies. Also, the orbitals with an energy higher than $15 \: \text{a.u.}$ are discarded to enhance the stability of the time propagation.

The numerical box radially extends to a size of $150 \: \text{a.u.}$ and is discretized with $1800$ nonuniformly distributed grid points with a mapping parameter of $\zeta = 1$ \cite{l.greenman10a}, from which we construct our FEM-DVR basis functions \cite{s.pabst14b}. The SES parameters are chosen as $r_0 = 20 \: \text{a.u.}$, $\theta = 40 ^\circ$, $\lambda = 1 \: \text{a.u.}$ such that the Hartree-Fock mean-field potential remains unscaled and that the continuum is rotated enough to expose the resonances. The maximum orbital angular momentum is $3$. For the time-dependent study, an initial state is propagated \cite{l.greenman10a} at a time step of $0.05 \: \text{a.u.}$ for a duration of $500 \: \text{a.u.}$ to give a sufficient frequency resolution. In the Gabor transformation, we use $\sigma = 2 \: \text{a.u.}$, which we select based on the spacing of the excitation energies for the two collective resonances. This choice represents an optimal trade-off between the spectral and temporal resolutions.

\subsection{\label{subsec:datases}Time-independent approach to Xe GDR:\\ SES method}
The diagonalization of the complex-scaled many-body Hamiltonian is achieved numerically by the iterative Arnoldi algorithm implemented in the \textsc{ARPACK} library \cite{d.sorensen12a, s.pabst14b, a.karamatskou13a}. An initial random vector is used to launch the iteration. Since we concentrate on the resonance modes in the linear-response regime, the eigenstates shown below are required to have a minimum overlap with the ground state through a dipole transition: $| \left( \Psi \right| \hat{D}_z |\Phi_0^{HF} ) | > 10^{-6}$, where $\hat{D}_z$ denotes the $z$ component of the dipole operator \footnote{The dipole operator in either the length or the velocity form can be used. The choice does not affect the Siegert energies. However, since the CIS theory is not strictly gauge-invariant \cite{n.rohringer06a}, the overlap $\left( \Psi \right| \hat{D}_z |\Phi_0^{HF} )$ has a slight degree of gauge dependence. In this study, we choose the velocity-form dipole operator, namely $\hat{P}_z$.} relative to the polarization direction of the electric field in a measurement.

Figure \ref{fig:ses_energy} shows the complex spectra of the energy eigenvalues for the case with only the intrachannel couplings and the case with both the inter- and intrachannel couplings, i.e., the full-model calculation within CIS. Ideally, the energy spectra in both cases should follow the structure predicted by the Balslev-Combes theorem \cite{e.balslev71a, n.moiseyev98b, y.ho83a, w.reinhardt82a}: the bound states remain on the real axis; the continuum is rotated clockwise by $2\theta$ degrees with respect to the $4d$ ionization threshold at $67.5\: \text{eV}$ \cite{a.thompson09a}; and the resonances are isolated above the continuum. However, the use of an incomplete basis set results in numerical artifacts such as branching of the continuum away from the threshold \cite{u.riss93a} and a rotation angle deviating from $2\theta$ \cite{o.alon92a}.

\begin{figure}
\includegraphics[angle=270]{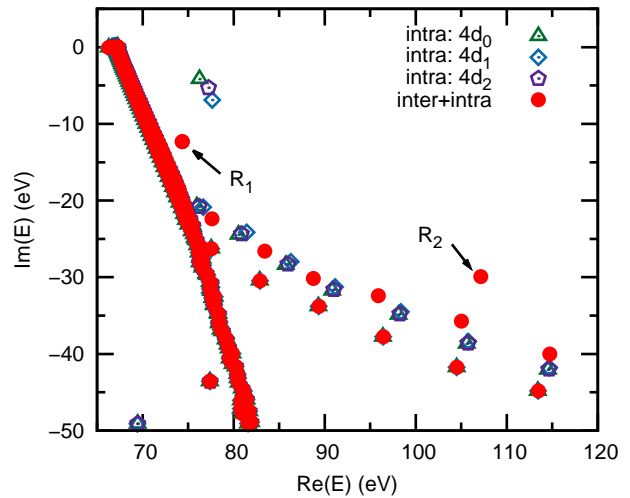}
\caption{\label{fig:ses_energy}(color online). Complex energy spectra of the complex-scaled many-body Hamiltonian for the intrachannel (``intra'') and full CIS (``inter+intra'') models. Horizontal and vertical axes represent the real and imaginary parts of the energy eigenvalues, respectively. Three close-lying resonances are found in the intrachannel calculation (indicated by hollow polygons); two far-separated resonances (labeled $R_1$ and $R_2$) are found in the full CIS calculation (indicated by solid dots).}
\end{figure}
 
First we focus on the result of the intrachannel-coupling model in Fig.~\ref{fig:ses_energy}. In this case, each eigenstate possesses a unique hole index $i$ and the contributions from different $4d$ ionization channels can be easily set apart. Three resonances are found, one for each $4d_{\pm m}$ channel. They lie fairly close to each other, forming a group of resonances around an energy with a real part $\approx 77 \: \text{eV}$ and an imaginary part $\approx -5.4 \: \text{eV}$, which corresponds to a lifetime $\approx 60 \: \text{as}$. The resonance positions and widths are detailed in Table \ref{tab:resonance}.

\begin{table*}[!]
\caption{\label{tab:resonance}Siegert energies of the resonances in the intrachannel model and the full CIS model. Results are listed for the SES method and for the Gabor analysis. For comparison, the predictions given by Wendin \cite{g.wendin73c} are included.}
\begin{ruledtabular}
\begin{tabular}{lcccccc}
& \multicolumn{2}{c}{SES \footnote{All SES values have an error bar of $0.1 \: \text{eV}$. This is calculated by varying over a reasonable range the numerical parameters such as the number of radial grid points, the maximum radial coordinate, and the SES parameters $\theta$ and $\lambda$.}} & \multicolumn{2}{c}{Gabor \footnote{In each calculation, the Gabor spectra are fitted numerically in a time interval $[t_i, t_f]$ at a time step of $3 \: \text{as}$. This gives the fitting parameters $\Xi_n, \Gamma_n$ as a function of time. The error bars are then defined as the standard deviations of $\Xi_n, \Gamma_n$ over the time sequence.}} & \multicolumn{2}{c}{Wendin \cite{g.wendin73c}} \\
 \cline{2-3} \cline{4-5} \cline{6-7}
 & $\Xi_n \: \text{(eV)}$ & $\Gamma_n  \: \text{(eV)}$ & $\Xi_n \: \text{(eV)}$ & $\Gamma_n  \: \text{(eV)}$ & $\Xi_n  \: \text{(eV)}$ & $\Gamma_n \: \text{(eV)}$\\
\hline
Intrachannel: $4d_0$ & $76.3$ & $8.3$ & $76.5 \pm 0.3$ \footnote{Using Eq.~(\ref{eq:gbsingle}) for one resonance in $[t_i, t_f] = [145 \: \text{as}, 363 \: \text{as}]$.} & $8.2 \pm 0.4$ \footnotemark[3] & -- & -- \\
Intrachannel: $4d_{\pm 1}$ & $77.6$ & $13.8$ & $77.9 \pm 0.7$ \footnotemark[3] & $13.5 \pm 0.4$ \footnotemark[3] & -- & -- \\
Intrachannel: $4d_{\pm 2}$ & $77.2$ & $10.6$ & $77.4 \pm 0.3$ \footnotemark[3] & $10.6 \pm 0.1$ \footnotemark[3] & -- & -- \\
Full ClS: $R_{1}$ & $74.3$ & $24.6$ & $80.4 \pm 0.7 $  \footnote{Using Eq.~(\ref{eq:gbdouble}) for two resonances in $[t_i, t_f] = [145 \: \text{as}, 169 \: \text{as}]$.}, $73 \pm 2$  \footnote{Using Eq.~(\ref{eq:gbsingle}) for one resonance in $[t_i, t_f] = [242 \: \text{as}, 363 \: \text{as}]$.} & $32 \pm 1$ \footnotemark[4], $17.8 \pm 0.4$ \footnotemark[5] & $74.3$ & -- \\
Full CIS: $R_{2}$ & $107.2$ & $59.9$ & $112 \pm 1$ \footnotemark[4] & $47 \pm 9$ \footnotemark[4] & $92.3$ & -- \\
\end{tabular}
\end{ruledtabular}
\end{table*}

In order to elucidate the origin of the small splitting among the intrachannel resonances, we perform another intrachannel calculation by approximating $\hat{H}_1$ with its monopole term \cite{s.pabst12b}. When so doing, the resonances associated with the $4d_0$, $4d_{\pm 1}$, and $4d_{\pm 2}$ channels have exactly the same resonance energy. Therefore, even without many-body effects, the electron excited from the different $4d_{\pm m}$ orbitals experiences different potentials owing to the shape of the non-spherical ionic core. This qualitative effect, although small, clearly exemplifies the impact of the ionic structure beyond the description of a simple spherically symmetric potential (e.g.~the Herman-Skillman \cite{f.herman63a} or Hartree-Fock \cite{a.szabo96a} potentials) or even an angular-momentum-dependent pseudopotential \cite{c.christiansen79a} widely used to model multielectron atoms in perturbative \cite{l.pi10a, j.cooper64a} or non-perturbative \cite{k.kulander88a, k.schafer93a, j.higuet11a} light fields.

Activating interchannel coupling, the group of intrachannel resonances splits into two resonance states distantly located in the complex energy plane in Fig.~\ref{fig:ses_energy}. We have checked carefully that their positions do not vary with the scaling parameters, so they are not numerical artifacts. In comparison with the intrachannel resonances, resonance $R_1$ in Fig.~\ref{fig:ses_energy} has almost the same excitation energy but a larger decay width; resonance $R_2$ is pushed away much further into the lower half of the complex energy plane. The splitting of the resonances highlights that many-body correlations are not just required for a quantitative agreement between theory and experiment \cite{j.cooper64a, u.fano68a, m.amusia67a, w.brandt67a, a.starace70a, a.zangwill80a}, but in fact give rise to fundamentally different resonance substructures underlying the GDR. Since Ref.~\cite{g.wendin73c} does not show calculations without many-body correlations, our study is the first to reveal the emergence of the collective resonances in the GDR from the intrachannel resonances.

With the interchannel interactions, each resonance cannot be attributed to a single ionization pathway. For both $R_1$ and $R_2$, the $4d_0^{-1}$, $4d_{\pm 1}^{-1}$, and $4d_{\pm 2}^{-1}$ hole populations \cite{l.greenman10a} have a rough ratio of $1:2:1$, which can be explained by an angular momentum analysis. Because the interchannel interactions strongly couple the various $4d_{\pm m}^{-1}$ hole states, it is crucial to consider the orbitals in addition to the one aligned along the polarization axis (i.e.~$4d_0^{-1}$) for the physical processes involving the GDR, e.g.~the giant enhancement in the HHG spectrum of xenon \cite{s.pabst13b}. We also compute the angular momentum composition of the excited electron, which shows a prominent $f$-wave character for both resonances. This is true in our intrachannel calculation, too. Indeed, the xenon GDR is dominated by the $4d \rightarrow \epsilon f$ transitions with roots in the independent-particle picture \cite{j.cooper64a, u.fano68a}.

Our CIS approach gives the total many-body resonance wave functions, which are not attainable using the plasma-type treatments including RPAE \cite{m.amusia67a, g.wendin71a, s.lundqvist80a}. We analyze $\left|\Phi_{1}\right)$ and $\left|\Phi_{2}\right)$ by decomposing them in terms of the orthonormal intrachannel basis set:
\begin{equation}
| \Phi_{n} ) = \sum a_{\pm m} | \Phi_{4d_{\pm m}} ) + | \Delta \Phi_{n} ), \: n = 1, 2. \label{eq:intradecomp}
\end{equation}
The first term in Eq.~(\ref{eq:intradecomp}) contains the projections onto the intrachannel resonance functions $| \Phi_{4d_{\pm m}} )$, and the second term symbolizes the remaining part with respect to the other intrachannel states. The complex weights $a_{\pm m}^2$ defined through the symmetric inner product are listed in Table \ref{tab:pop}. For both $R_1$ and $R_2$, $a_0^2$, $a_{\pm 1}^2$, and $a_{\pm 2}^2$ have the same order of magnitude. For $R_1$, the intrachannel resonances account for a weight of $\sum_m a_m^2 \approx  0.7$ out of a total norm of $1$; for $R_2$, they contribute a weight $\approx 0.3$. This means that the interchannel interactions do not only mix all the intrachannel resonance states, but also mix in continuum states to form the new resonances. At this stage, we clearly see in our CI language why we can refer to $R_1$ and $R_2$ as new collective dipolar modes: they are entangled particle-hole states involving the various $4d^{-1}_{\pm m}$ hole states and do not resemble any single intrachannel resonance wave function. In RPAE, collective excitations are not directly defined by the many-body wave functions themselves, but by a coherent sum over different particle-hole states in evaluating the dipole \cite{m.amusia00a} or dielectric function \cite{g.wendin73c, m.amusia74a} matrix elements. Note that both collective states have a significant overlap with the Hartree-Fock ground state via a dipole transition, with $\left( \Phi_1 \right| \hat{D}_z |\Phi_0^{HF} ) \approx 1.1$ and $\left( \Phi_2 \right| \hat{D}_z |\Phi_0^{HF} ) \approx 1.9$ for $R_1$ and $R_2$, respectively.

\begin{table}
\caption{\label{tab:pop}The complex weights $a_{\pm m}^2$ with respect to the intrachannel resonance states $| \Phi_{4d_{\pm m}} )$ for the collective resonances $\left|\Phi_{1}\right)$ and $\left|\Phi_{2}\right)$ in the SES calculations.}
\begin{ruledtabular}
\begin{tabular}{lcc} 
& $R_1$ & $R_2$ \\
\hline
$a_0^2$ & $0.175 - 0.033i$ & $0.075 + 0.042i$ \\
$a_{\pm 1}^2$ & $0.345 + 0.009i$ & $0.106 + 0.003i$ \\
$a_{\pm 2}^2$ & $0.198 -  0.018i$ & $0.081 + 0.028i$\\
$\sum_m a_m^2$ & $0.719 - 0.042i$ & $0.262 + 0.073i$
\end{tabular}
\end{ruledtabular}
\end{table} 

The positions and widths of the resonances in the full model can be found in Table \ref{tab:resonance}. The resonance positions calculated by Wendin \cite{g.wendin73c} are also listed for comparison. The resonance position of $R_1$ agrees perfectly with that given by Wendin. The position for $R_2$ differs from his number by $15 \: \text{eV}$, but both positions are compatible with the spectral blur observed in the experimental cross section.

The most likely reason for the discrepancy of $\Xi_2$ predicted by Wendin and our result is the approximate condition Wendin used to find collective excitations from his effective dielectric function. In principle, a collective resonance corresponds to a complex frequency where both the real and imaginary parts of the many-body dielectric function simultaneously become zero \cite{n.march95a, m.amusia74a}. An estimated resonance position can be found by determining along the real energy axis a root for the real part of the dielectric function, \textit{provided the damping of the true resonance is sufficiently small} \cite{m.amusia74a}. As one is dealing with two rather broad resonances in the case of the xenon GDR (particularly for $R_2$), this approximate condition, which is adopted by Wendin \cite{g.wendin73c}, is not strictly applicable. In the Appendix\ref{app:dielec}, we demonstrate how this simplified condition of finding the zeros of the dielectric function can result in Siegert energies that deviate substantially from the true resonance poles. Based on this argument, the Siegert energies provided by the present study are most likely to be more reliable.

\subsection{\label{subsec:datagb}Time-dependent approach to Xe GDR:\\ Gabor analysis of autocorrelation functions}
To probe the resonances associated with one-photon absorption, the initial wave packet can be conveniently set as
\begin{equation}
\left.| \Psi(0) \right) = \hat{D}_z \left. |\Phi_0^{HF} \right). \label{eq:initial}
\end{equation}
This is equivalent to creating a wave packet via a delta-kick pulse polarized along the $z$ axis. Since it is well known that the xenon GDR exhausts all the oscillator strength in the XUV \cite{m.amusia00a, c.brechignac94a}, the wave packet prepared in this way is largely composed of the relevant resonances, and its autocorrelation function $C(t)$ is expected to take the form assumed in Eq.~(\ref{eq:auto}).

\begin{figure}
\includegraphics[angle=270]{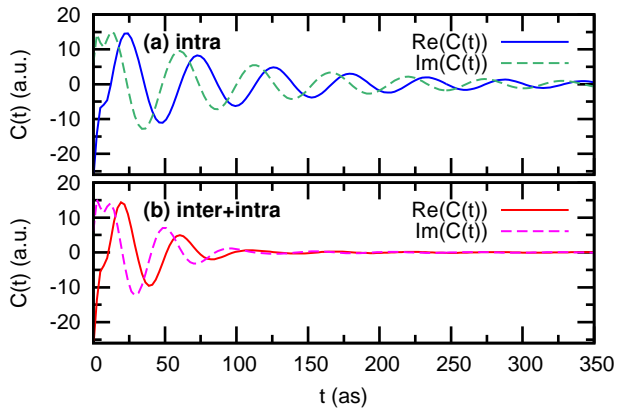}
\caption{\label{fig:corr}(color online). Autocorrelation functions $C(t)$ as a function of time for (a) the intrachannel model (``intra'') and (b) the full CIS model (``inter+intra''). Solid curves for the real part and dashed curves for the imaginary part of $C(t)$.}
\end{figure}

The wave packet subsequently undergoes field-free relaxation. Fig.~\ref{fig:corr} plots the time evolution of the complex-valued auto-correlation function for both the intrachannel and full CIS models. The raw data in both cases look like a simple damped oscillator without much structure apart from some spikes in the beginning. This suggests that there are some dynamics that rapidly disappear. Including the interchannel interactions causes the autocorrelation function to attenuate faster and to ring at a higher frequency.

\begin{figure}
\includegraphics[angle=270]{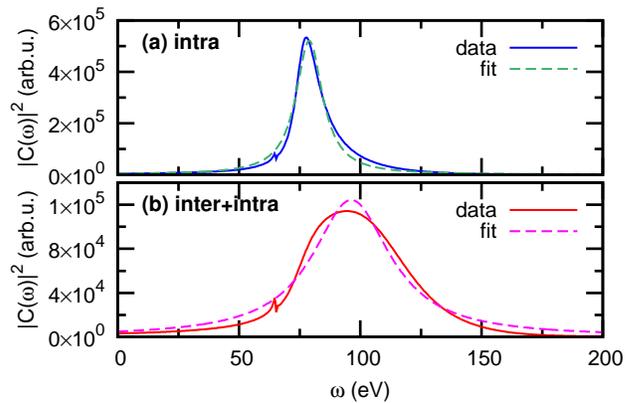}
\caption{\label{fig:ft}(color online). Autocorrelation functions $|C(\omega)|^2$ in frequency domain for (a) the intrachannel model (``intra'') and (b) the full CIS model (``inter+intra''). Solid curves for the data and dashed curves for the fit with Eq.~(\ref{eq:ftsingle}).}
\end{figure}

The above features are more pronounced if we look at the auto-correlation functions in the frequency domain as illustrated in Fig.~\ref{fig:ft} \footnote{At this step, $C(t)$ is exponentially damped by hand to reduce the spectral aliasing error. More specifically, $C'(t) \coloneqq C(t) e^{-t/\tau}$ with $\tau \approx 108 \: \text{a.u.}$ \cite{d.krebs13a}. This globally adds a small lifetime to all the wave-packet components and can be easily rectified afterwards.}. The Fourier transform $|C(\omega)|^2$ in each calculation shows one smooth peak, accompanied by Rydberg series preceding the $4d$ ionization energy \cite{d.krebs14a}. The linewidth of the Rydberg states is narrow, since a $4d^{-1}_{\pm m}$ hole decays on the femtosecond time scale \cite{p.lablanquie02a}. Although the SES method yields three intrachannel resonances, they cannot be distinguished and really act as a group here. Switching on the interchannel couplings broadens and weakens the peak as well as displaces it to a higher frequency, similar to what is seen in the photoabsorption spectra \cite{d.krebs14a}. In the full model, the two resonances given by the SES analysis also cannot be resolved in the Fourier domain.
 
We can extract the effective Siegert energy for the single peak in Figs.~\ref{fig:ft}(a) and \ref{fig:ft}(b). The data are fitted numerically with Eq.~(\ref{eq:ftsingle}) utilizing the nonlinear least-squares Marquardt-Levenberg algorithm. This yields $(\Xi, \Gamma) = (79.0 \: \text{eV}, 12.9 \: \text{eV})$ for the intrachannel model and $(96.3 \: \text{eV}, 38.9 \: \text{eV})$ for the full CIS model. Notice that $|C(\omega)|^2$ in the full CIS model is relatively poorly described by its Lorentzian fit and is more asymmetric, a hint to the multiple resonances behind the huge spectral hump of the xenon GDR.

\begin{figure}
\includegraphics[angle=270]{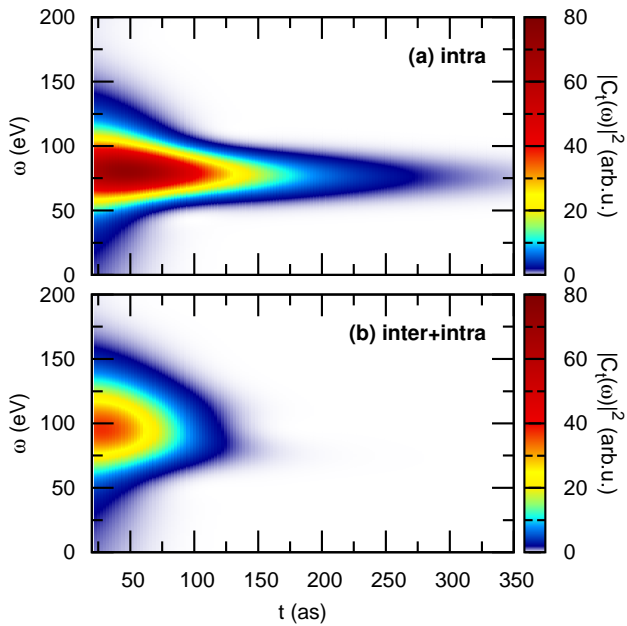}
\caption{\label{fig:gb_3d}(color online). Autocorrelation functions $|C_t(\omega)|^2$ in the combined time-frequency domain for (a) the intrachannel model (``intra'') and (b) the full CIS model (``inter+intra'').}
\end{figure}

Now, we are in a position to go beyond the standard spectral method and to investigate the xenon GDR in the combined time-frequency domain. Fig.~\ref{fig:gb_3d} depicts the Gabor transform $|C_t(\omega)|^2$ for both the intrachannel and full CIS models \footnote{At this point, we filter out the tiny contribution from the Rydberg series to focus on the properties pertaining to the GDR. Another autocorrelation function $C''(t)$ is defined by removing the Rydberg series in $C'(\omega)$, interpolating the remaining curve using the natural cubic splines, and then inverse Fourier transforming it back to the time domain.}. In addition to the information that is already revealed by $C(t)$ and $|C(\omega)|^2$, one salient new feature emerges: the spectral distribution in the intrachannel case dies out almost symmetrically over time, whereas the spectral distribution in the full CIS model decays asymmetrically, with the maximum shifting to a lower energy. In the vicinity of $150 \: \text{as}$, one can clearly recognize two frequency components in Fig.~\ref{fig:gb_3d}(b), and can deduce that the higher-energy one has a shorter lifetime.

\begin{figure}
\includegraphics[angle=270]{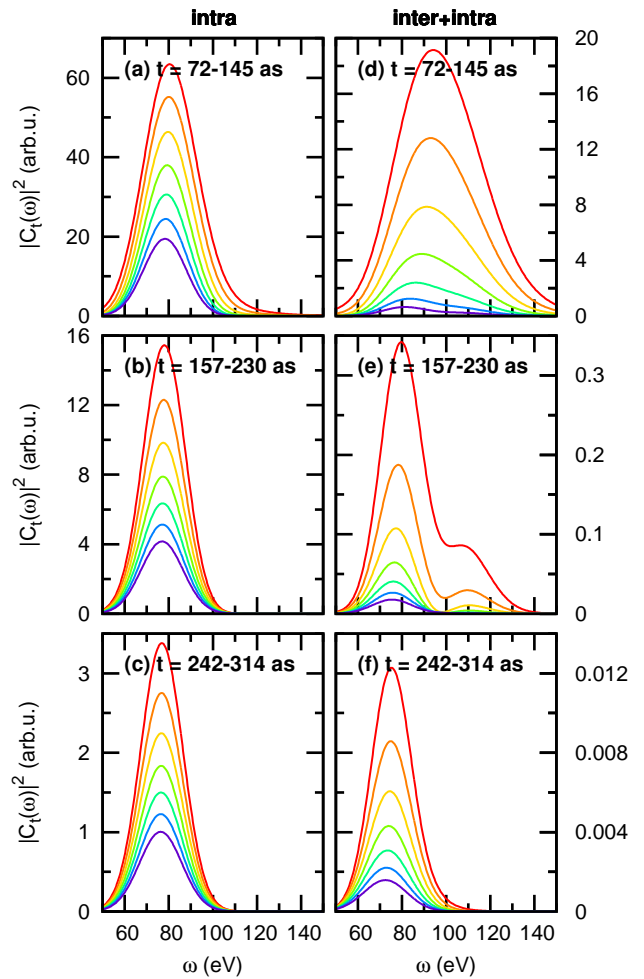}
\caption{\label{fig:gb_2d}(color online). Decay dynamics of the Gabor spectra $|C_t(\omega)|^2$ at consecutive time intervals. Panels (a)(b)(c) in the the left column for the intrachannel model (``intra''); panels (d)(e)(f) in the right column for the full CIS model (``inter+intra''). In each panel, the time step between two neighboring lines is about $3 \: \text{as}$.} 
\end{figure}

Figure \ref{fig:gb_2d} presents snapshots of the frequency distributions $|C_t(\omega)|^2$ in Fig.~\ref{fig:gb_3d} at consecutive time steps, where the characteristics we allude to can be even better visualized. In the intrachannel model, Figs.~\ref{fig:gb_2d}(a)(b)(c) exhibit one single, decaying peak around $80 \: \text{eV}$, which arises from the group of three intrachannel resonances in the SES calculation. Upon closer examination, we find that the peak position gradually moves to a slightly lower frequency. This is in accordance with the SES study that the lowest-lying intrachannal resonance has the smallest decay width (see Table \ref{tab:resonance}).

In the full CIS model, the time evolution of the transient spectral distribution is fundamentally different. In Fig.~\ref{fig:gb_2d}(d), the initial spectrum displays a broad and nearly symmetric peak located around $90 \: \text{eV}$. However, the spectrum soon becomes highly asymmetric with the maximum shifting to a lower energy, and decays faster than the intrachannel case. The successive dynamics in Fig.~\ref{fig:gb_2d}(e) vividly picture how the substructures in the GDR develop---two peaks can be identified, and the higher-lying one fades away much quicker. This is followed by Fig.~\ref{fig:gb_2d}(f), where the higher-lying mode has completely vanished and only the lower-lying one remains, with a position similar to that of the intrachannel resonance. Just based on simple observations, the Gabor analysis intuitively illuminates how the interchannel interactions result in the damping and fragmentation of the resonances, as well as a rough idea of the resonance positions and widths. Benefitting from the fact that different spectral components have different lifetimes, the Gabor analysis successfully disentangles the two fundamental collective modes that cannot be separated by the Fourier analysis.

Next, the Siegert energies are quantitatively determined following the logic presented in Sec.~\ref{subsec:gb}. Based on the \textit{a priori} input from the SES study, we perform the analysis in the intrachannel case for each $4d_{\pm m}$ ionization channel. One peak in the Gabor profiles at subsequent time steps is then fitted numerically with Eq.~(\ref{eq:gbsingle}). The outcomes are tabulated in Table \ref{tab:resonance}, and are in excellent agreement with the SES results within the error bars. Particularly, the Gabor analysis captures the minute splitting trend of the resonance energies, i.e.~$\Xi_{4d_{\pm 1}} > \Xi_{4d_{\pm 2}}> \Xi_{4d_0}$ and $\Gamma_{4d_{\pm 1}} > \Gamma_{4d_{\pm 2}}> \Gamma_{4d_0}$.

For the full CIS model, the fitting process is dissected into two stages. At the first stage, roughly corresponding to the time interval shown in Fig.~\ref{fig:gb_2d}(e), two resonances are singled out. Resorting to Eq.~(\ref{eq:gbauto}) with $n = 2$, the Gabor spectrum has the approximate analytical expression:
\begin{align}
\left| C_t^{(2)}(\omega) \right|^2
& \approx f_{1,t}(\omega) + f_{2,t}(\omega) \nonumber \\
& + 2 \left[ f_{1,t}(\omega)f_{2,t}(\omega)\right]^{\frac{1}{2}} f_{1-2,t}(\omega),
\label{eq:gbdouble}
\end{align}
where
\begin{align}
f_{n,t}(\omega) \coloneqq & \left| a_n^2 \right|^2 e^{\frac{\sigma^2 \Gamma_n^2}{4}} e^{-\Gamma_n t} e^{-\sigma^2(\omega - \Xi_n)^2}, \: n = 1, 2 \nonumber \\
f_{1-2,t}(\omega) \coloneqq & \cos \left[ \frac{\sigma^2 (\Gamma_2 - \Gamma_1) \omega}{2} + (\Gamma_2 - \Gamma_1)t \right. \nonumber \\
 & - \left. \frac{\sigma^2 (\Gamma_2 \Xi_2 - \Gamma_1 \Xi_1)}{2} + ( \phi_2-\phi_1) \right].
\label{eq:gbdoubledef}
\end{align}
In Eq.~(\ref{eq:gbdouble}), the first two terms are the individual contributions from $R_1$ and $R_2$, and the third one is their interference. In Eq.~(\ref{eq:gbdoubledef}), $\phi_n$ denotes the phase of $a_n^2$. The data are fitted with the above formulae, and the Siegert energies are given in Table \ref{tab:resonance}. The error bars here are bigger than those in the intrachannel case, and the extracted Siegert energies deviate from the SES ones, especially for the decay width of the faster-decaying $R_2$. This is possibly due to the increasing difficulties in the nonlinear fitting procedure (particularly from the interference term). Also, in order to arrive at the analytical expression for $|C_t(\omega)|^2$, Eq.~(\ref{eq:auto}) assumes that all the contributions to $C(t)$ from the bound states and the continuum can be neglected, which works worse if the resonances decay relatively fast.

At the second stage, which nearly coincides with the time interval shown in Fig.~\ref{fig:gb_2d}(f), only one resonance is seen. Using Eq.~(\ref{eq:gbsingle}), we produce another Siegert energy for $R_1$ in Table \ref{tab:resonance}. It does not fully agree with that retrieved at the former stage, especially in terms of the resonance width. The two-resonance model used at the first stage is only applicable in a short period of time, where the resonance parameters certainly cannot fluctuate too much. Hence, this discrepancy reflects further numerical instability in the fitting parameters that cannot be entirely represented by the previously calculated uncertainties. The Siegert energy acquired at this second stage seems closer to the SES one. Nevertheless, as the Gabor spectrum in this time interval has a fairly weak amplitude, the contribution from the Rydberg series (after the filtering) inevitably kicks in, which lowers the effective energy position and width for $R_1$.

The most appealing feature of the Gabor analysis is that it provides an intuitive dynamical view on the competition between various spectral components, which may be connected to what is measured in a pump-probe experiment \cite{a.wirth11a, a.kaldun14a, e.power10a}. However, it is apparent that the Gabor method cannot quantify Siegert energies as accurately as the SES approach. Considering the deviations from the SES results, the energy uncertainties, and the discrepancy between the resonance energies for $R_1$ extracted at two different stages, the Gabor analysis gives an overall energy resolution $\approx 10 \: \text{eV}$.

\subsection{\label{subsec:5s5p}Influence of  5s and 5p orbitals}
In the above Secs.~\ref{subsec:datases} and \ref{subsec:datagb}, the active ionization channels lie in the $4d$ shell; the outer $5s$ and $5p$ shells are frozen. This is an assumption made in Ref.~\cite{g.wendin73c} as well.

Our SES calculations show that including the $5s$ and $5p_{\pm m}$ channels leads to no qualitative but only minor quantitative modifications to the previous discussion. Hence, the xenon GDR mainly stems from the many-body correlations involving the ten $4d$ electrons \cite{m.amusia00a}, and the $5s$ and $5p$ electrons are only small admixtures. The Siegert energies for the three intrachannel resonances remain the same as in Table \ref{tab:resonance}. In the full CIS model with active $4d$, $5s$ and $5p$ shells, the Siegert energies are slightly revised to $(\Xi_1, \Gamma_1) = (73.4 \: \text{eV} , 24.7 \: \text{eV})$ and $(\Xi_2, \Gamma_2) =(111.8 \: \text{eV} , 58.2 \: \text{eV})$. Note that the higher-lying $R_2$ is more sensitive to the effects of the outer shells.

\section{\label{sec:conclude}Conclusion}
In this paper, we disentangle two fundamental collective dipolar resonances that cannot be resolved in the photoabsorption cross sections associated with the xenon GDR. In extension of Wendin's pioneering work \cite{g.wendin71a}, we achieve a complete theoretical characterization of the resonance substructures by two complementary methods within the framework of the CIS theory. It is very likely that the Siegert energies given by the current study are more accurate than those given by Wendin, as our methodology for finding the resonance poles is not limited to weakly damped oscillations. The time-independent SES approach demonstrated here is the first example of treating collective resonances in multielectron atoms with the complex scaling technique. The time-dependent Gabor analysis extends the standard Fourier analysis to the combined time-frequency domain, such that strongly overlapping resonances living on different time scales can be more easily separated.

Our work provides a deeper insight into the nature of the GDR: the group of three close-lying intrachannel resonances splits into two far-separated resonances upon the inclusion of interchannel couplings primarily involving the $4d$ electrons. The two resonances are new collective modes in the sense that they must be written as a superposition of various particle-hole wave functions. When the excited electron is still near the ion, a strongly entangled particle-hole pair is formed. This leads to the strong mixing of the various $4d_{\pm m}^{-1}$ ionic states, the entire $4d$ shell thus exhibiting collective plasma-like oscillations as a whole.

We specify the Siegert energies for the two collective resonances. However, the exact values need further theoretical refinement. The CIS theory only contains one-particle--one-hole configurations (in addition to the Hartree-Fock ground state). Hence, real and virtual double excitations \cite{a.starace82a, m.amusia00a, g.wendin73c} are among the physical processes outside the scope of the current study. Nonetheless, since TDCIS (in the velocity gauge) produces a peak position in good agreement with the experimental photoabsorption cross section \cite{d.krebs14a}, we expect that inclusion of double excitations would not affect the resonance parameters substantially. 

Finally, we note that a recent experiment at the free-electron laser FLASH, using an XUV nonlinear spectroscopy technique, has provided the first direct evidence for the two collective dipolar resonance states associated with the xenon GDR \cite{t.mazza14a}. Thus, it may be expected that experiments of this type will provide an opportunity to test the predictions presented in this paper.

\begin{acknowledgments}
This work was supported in part by the Deutsche Forschungsgemeinschaft under Grant No. SFB 925/A5.
\end{acknowledgments}

\appendix*
\section{\label{app:dielec} Consequence of using the approximate condition to find zeros of dielectric functions}
As briefly explained at the end of Sec.~\ref{subsec:datases}, the approximate condition invoked by Wendin \cite{g.wendin73c} to find zeros of the many-body dielectric function can result in Siegert energies that substantially deviate from the true resonance poles. In this section, we provide numerical evidence in support of the above statement.
\begin{figure}
\includegraphics[angle=270]{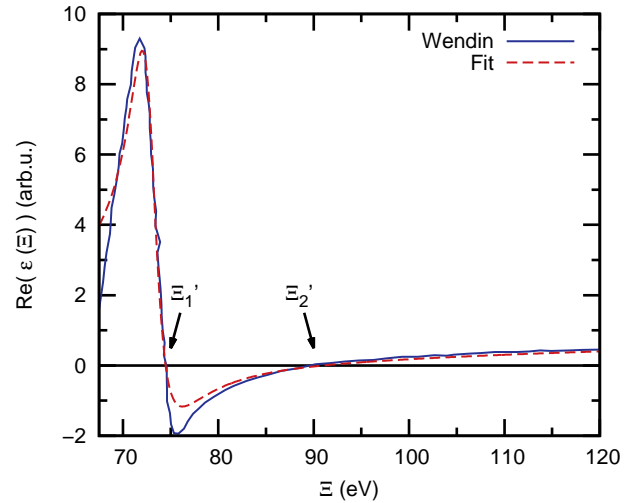}
\caption{\label{fig:dielec}(color online). Real part of the many-body dielectric function $\operatorname{Re} ( \epsilon$) along the real energy axis $\Xi$. Solid curve for Wendin's data \cite{g.wendin73c} and dashed curve for the fit with Eq.~(\ref{eq:dielectric}). The two resonance positions determined in the spirit of Wendin's work are labeled $\Xi_1 '$ and $\Xi_2 '$.}
\end{figure}

A collective resonance obtained from diagonalizing the complex-scaled many-body Hamiltonian corresponds to an energy $E_n = \Xi_n - i \Gamma_n / 2$ in the complex energy plane where both the real and imaginary parts of the many-body dielectric function $\epsilon (E)$ simultaneously vanish \cite{n.march95a, m.amusia74a}. In the limit of $\Gamma_n \rightarrow 0$, this exact condition is reduced to finding the roots of the real part of $\epsilon (\Xi)$ along the real energy axis $\Xi, \: \Gamma = 0$ \cite{m.amusia74a}.

For a dielectric function with two poles in the energy range of interest, it must follow the simple analytical structure \cite{n.march95a}:
\begin{equation}
\epsilon (E) = \frac{1}{1-\left( \frac{a_1}{E - E_1} +\frac{b_2}{E - E_2}  \right)}, \label{eq:dielectric}
\end{equation}
where $a_1$ and $a_2$ are two complex numbers.

To extract Wendin's dielectric function for the xenon GDR, we fit the real part of his data with $\operatorname{Re} (\epsilon (\Xi))$ in Eq.~(\ref{eq:dielectric}). The $a_n$ are treated as fitting parameters, and the $E_n$ as known constants with the values given by the SES method in Table \ref{tab:resonance} (as mentioned in Sec.~\ref{subsec:datases}, we believe that our Siegert energies are closer to the true ones). The real part of the reconstructed dielectric function (with $a_1 = -10.0 - 19.7 i$ and $a_2 = -1.6 + 6.4 i$) is shown in Fig.~\ref{fig:dielec}, and nicely describes Wendin's result.

The real part of $\epsilon (\Xi)$ we retrieve passes through the real energy axis at $\Xi_1 '= 74.5 \: \text{eV}$ and $\Xi_2 ' = 91.0 \: \text{eV}$, which do not coincide with the excitation energies of the true collective poles $\Xi_1 = 74.3 \: \text{eV}$ and $\Xi_2 = 107.2 \: \text{eV}$. This strongly indicates that the approximate condition of finding the roots of $\operatorname{Re} (\epsilon (\Xi))$ as Wendin did does not suffice to provide accurate predictions for the short-lived collective resonances. In particular, the estimated $\Xi_2$ for the shorter-lived $R_2$ amounts to an error of $16.2 \: \text{eV}$. In our opinion, the most consequential approximation Wendin made does not lie in the way he constructed the dielectric function, but in the way he searched for the collective poles, which is suitable only for weakly damped oscillations \cite{m.amusia74a}.

\end{document}